\documentclass[aip,prb,amsmath,amssymb,reprint]{revtex4}
\usepackage[dvips]{graphicx}
\usepackage{dcolumn}
\usepackage{color}
\usepackage{ulem}
\usepackage{mathptmx, helvet, courier}
\usepackage{bm,amsmath,amssymb}
\usepackage{threeparttable}
\usepackage{multirow}
\usepackage{array}
\usepackage{etoolbox}

\begin{document}

\title{Proteins at air-water and oil-water interfaces in an all-atom model}

\author{Yani Zhao}
\affiliation{Institute of Physics, Polish Academy of Sciences,
 Al. Lotnik{\'o}w 32/46, 02-668 Warsaw, \\ Poland}

\author{Marek Cieplak}
\email{mc@ifpan.edu.pl}
\affiliation{Institute of Physics, Polish Academy of Sciences,
 Al. Lotnik{\'o}w 32/46, 02-668 Warsaw, \\ Poland}

\date{\today}

\begin{abstract}
\noindent
We study the behavior of five proteins at the
air-water and oil-water interfaces by  
all-atom molecular dynamics. The proteins are found to
get distorted when pinned to the interface. This behavior is 
consistent with the phenomenological way of introducing
the interfaces in a coarse-grained model through a
force that depends on the hydropathy indices of the residues.
Proteins couple to the oil-water interface stronger
than to the air-water one. They diffuse slower at 
the oil-water interface but do not depin from it, whereas
depinning events are observed at the other interface.
The reduction of the disulfide bonds slows the diffusion down. 
\end{abstract}

\maketitle

\section{Introduction}

Fluid-fluid interfaces, such as oil-water, air-water, and vapor-water 
systems, are fast changing and rough at the molecular time and length
scales. 
Despite these transient features, the interfaces exhibit preferred
orientations of the water molecules as evidenced by the surface
vibrational sum frequency spectroscopy (VSFS) \cite{richmond2,oilwater}.
They have also been a subject of theoretical research. For instance,
Sega and Dellago \cite{sega2017} have studied differences in the density
profile and surface tension between seven models of the molecules
of water near the air-water interface.\\

Proteins staying at the fluid-fluid interfaces add an additional
complexity to this dynamics because of the variety of the chemical
groups in the side chains that interact with molecules of the fluids
in different manners. In bulk water, the folded chain of a globular
protein is stabilized by hydrogen bonds and by
interactions with the molecules of water.
However, if there is a nearby interface, the protein may
diffuse in its direction and get adsorbed. 
Polar and charged groups are then drawn into the
bulk water, whereas the hydrocarbon groups seek the oil or vapor.
As a result, the protein gets distorted and may loose its 
biological functionality \cite{Adam,Norde,Graham}.
Many proteins coming to the interface may form a viscoelastic film
\cite{Murray,Cicuta2007,Wierenga2006,Leheny,Leheny1,Leheny2,Sollich1997,Malcolm2006}. 
Understanding the behavior of protein films is important for the
food industry, production of cosmetics, and physiology.\\

A theoretical description of such large dynamical systems requires making
simplifications. In an approach proposed in ref. \cite{airwater}, the
protein at the air-water interface is described by a coarse-grained 
structure-based model \cite{JPCM}, the solvent is implicit, and the
effect of the interface is introduced through a force that is
coupled to the hydropathy indices, $q_i$, associated with the 
amino acid residues. There are many hydropathy scales \cite{Palliser} and
the model makes use of the one proposed by Kyte and Doolittle\cite{Kyte}.
In this scale, the values of $q_i$ range between --4.5 for the
polar and charged ARG and 4.5 for the most hydrophobic ILE.
The interface-related force acting on the $i$th ${\alpha}$-C
atom is given by \cite{airwater}
\begin{equation}
F_i^{wa}=q_i\;A \;\frac{\text{exp}(-z_i^2/2W^2)}{\sqrt{2\pi}W}
\label{hyforce}
\end{equation}
where the amplitude $A$ is set equal to 10$\;\epsilon$, and $W$=5 {\AA}.
Parameter $W$ specifies the width of the interface and parameter
$A$ has been selected so that when a protein comes to the
interface it does not depin from it. Smaller values of $A$ allow for
depinning of several proteins that have been studied \cite{airwater}. 
Parameter $A$ is given in terms of $\epsilon$
which is equal to the depth of the potential in the effective
contact interaction between two residues. $\epsilon$ has been
calibrated \cite{plos} to be of order 110 pN\;{\AA}.\\


This  phenomenological model has been used to demonstrate glassy
behavior of films made of lysozymes, protein G, and hydrophobin \cite{airwater}.
Subsequently, it was employed to study interface-induced 
topological changes in proteins \cite{topology}, such as tying
and untying of a shallow knots in the proteins. 
Most recently, it was used to explain stabilization
of beer foams by adsorption of the isoforms of the 
lipid transfer protein 1 (LTP1) to
the surfaces of CO$^2$ bubbles \cite{beer}.\\

Here, we focus on the atomic-level justification of the model -- the subject
that we started to discuss in the context of the beer proteins \cite{beer}.
We perform all-atom simulations for five proteins: tryptophan cage (PDB:1L2Y),
streptococcal protein G (PDB:1GB1), hydrophobin HFBI (PDB:2FZ6),
LTP1 (PDB:1LIP), and the hen egg-white
lysozyme (PDB:2LYZ). The sequential lengths of these proteins
are correspondingly 20, 56, 72, 91 and 129. For brevity, we shall
refer to the proteins by using their PDB structure codes.
Three of these proteins, 1LIP, 2FZ6 and 2LYZ, are stabilized by
$n_{SS}$ disulfide bonds. For the three proteins, $n_{SS}$ is
equal to 4, but we also consider the reduced systems in which $n_{SS}$=0.
We study single proteins and not films made of the proteins.
The overall hydropathy of a protein with the  sequential
length $N$ can be characterized by
the parameter $H=1/N\sum_{i=1}^{N}q_i$. For the hydropathy scale
of Kyte and Doolittle, it is equal to
--0.95, --0.63, +0.27, --0.38, and --0.47 for 1L2Y, 1GB1,
2FZ6, 1LIP and 2LYZ, respectively. Thus, only 2FZ6 (hydrophobin)
is overall hydrophobic whereas the remaining proteins
are overall hydrophilic.  The reason to select these proteins is that 
they have already been studied within the coarse-grained model either in 
ref.~\cite{airwater} (1GB1, 2FZ6, and 2LYZ) or in ref.~\cite{beer} (1LIP).  
2LYZ is one of the proteins used
in the experimental studies of thin films \cite{Leheny2}.\\

The phenomenological model predicts that the air-water interface
distorts proteins and leads to adjustment of their orientation
to optimize the hydropathy-related forces. Here, we demonstrate
that the all-atom model acts in a similar fashion, but the proteins
do not necessarily stay pinned at the interface. Permanent pinning
at the air-water interface 
occurs for 1L2Y and 2FZ6,  but not for the remaining proteins.
In the latter case, we study multiple pinning events and determine
their durations. Finally, we extend the studies to proteins
at the oil-water interface. In this case, all of the proteins
studied stay pinned there. \\

The formation of interface between two immiscible phases
is a common phenomenon observed in ordinary everyday 
events (like beer foams \cite{beer} ) as well as in biological processes. The 
understanding of the molecular processes at
interfaces between liquids with 
different hydrophobicities is relevant to the biological activity of 
proteins in living organisms, and is important to the development of
new systems through the tools of biotechnology.
For example, the  blood/biomaterial interfaces can be well 
mimicked by a hydrophobic/hydrophilic interfacial model \cite{tripp}. The 
previous experimental work \cite{tripp} has been exhibited that, proteins with 
hydrophobic surfaces possess fast adsorption rates at the air-water interface, 
while these with hydrophilic surfaces exhibit low affinity at the interface. Our 
simulating results coincide very well with the experimental observation, 
i. e. the affinity of 2FZ6 at the air-water interface is much 
higher than that of 1GB1. 

\section{Methods}

The simulations are performed by using the NAMD \cite{NAMD} all-atom molecular 
dynamics
package with the CHARMM22 force field \cite{charmm2}.
The TIP3P water model \cite{tip3p} is used.  The system is placed
in a simulation box which extends between --$L_x$ and $L_x$ 
in the $x$ and $y$ directions, and  --$L_z$
and $L_z$ along the $z$-direction. We take $L_x$=50 {\AA} and $L_z$
is equal to 100 {\AA} for the air-water interface and 50 {\AA} for
the oil-water one. In both cases, the water molecules are placed
initially in the space corresponding to $z\le 0$, 
and the interface is extended in the $x-y$ plane.
In order to make the water molecules prefer staying in the
lower half of the box, we place a hydrophilic wall at the  
bottom of the air-water system (at $z$=--$L_z$). 
The presence of the wall efficiently prevents the diffusion of water 
molecules to the air phase (with $z>0$) under the 
periodic boundary conditions, which are applied to all directions for both 
interfaces. However, this wall is not
needed in the oil-water case. The immiscible nature of oil and water  molecules
generate an obvious interface between the two phases.
The wall is composed of 6728 asparagines \cite{beer}
that are arranged into a single layer. The 
$\alpha$-C atoms are anchored to the sites of the 
[001] face of an fcc lattice with the lattice constant of 5 {\AA}. 
The side groups of the asparagines are directed into the box
and they stay frozen during the simulations.\\

The major components of olive oil is oleic acid 
which constitutes 55 to 79\% of the fluid \cite{oliveoil}. 
The molecule is a chain of 18 carbon atoms 
(the length of the carbon-carbon bonds ranges from 1.31 to 1.53 {\AA})
with the first one forming bonds with two oxygen atoms as
illustrated in Fig.~S1 in the Supplementary Information (SI).
The density of oleic acid is 0.89 g/cm$^3$. \cite{pubchem}
To generate the oil-water interface, we place the oleic acid molecules
initially in the upper half of the box.\\

It is difficult to perform full-scale all-atom simulation in which
the protein is placed in the bulk water, gets equilibrated, and then diffuses
to the interface. Such a simulation would require tremendous averaging
necessitated, by the dynamic nature of the interface, evolution of
the protein, and the motion of the water molecules in the bulk.
Therefore, we perform a simplified calculation in which we just
explore the effects of the interface on the protein.
In the initial state, we place the protein right at the interface
so that the center of mass of the protein is located at (0,0,0).
The protein can have various orientations that can be characterized
by the direction that the hydropathy vector makes with the $z$-axis.
The vector is defined \cite{airwater} as
$\vec{h} = \frac{1}{N}
\sum_{i=1}^{N}q_i \vec{\delta_i}$. 
$\vec{\delta_i}$ is the position vector of the $i$th 
residue with respect to the center of mass
of the protein, and $N$ is the total number of the residues.
We have focused on two orientations: in orientation I
the native $\vec{h}$ points in the positive $z$-direction
and in orientation II, in the opposite direction.
Orientation I is expected to provide a preferred alignment.\\

To neutralize the charge of the protein, Na$^+$ or Cl$^-$ ions are added 
to the system. For example, four Na$^+$ ions, two Cl$^-$ ions, one 
Cl$^-$ ion and eight Cl$^-$ were required in the case of
1GB1, 1LIP, 1L2Y, and 2LYZ, respectively. 
For 2FZ6, no nuetralizing ions are needed.
The electrostatic
interactions are accounted for by using the Particle Mesh 
Ewald method \cite{Darden}. \\

In the next stage, the interfacial system is minimized 
for 0.02 ns by the conjugate gradient algorithm with all atoms being allowed to 
move. The minimization is performed at constant volume.
The equilibration of the system is performed in two steps:
2 ns at $T$=150 K and then $\ge$10 ns at 300 K. 
In both steps, the system is simulated 
in the canonical ensemble (NVT) with the Langevin temperature control. 
The timestep is 2 fs in our simulations. Each trajectory
lasts for 50 ns for proteins 1L2Y and 1LIP at both air-water and oil-water
cases. It lasts for 10 ns for other proteins.\\

The instantaneous orientation of the protein is characterized by a  
parameter $\theta$, which is the angle formed between the 
instanataneous hydropathy vector $\vec{h}$ and the positive $z$-axis.
$\theta$ is initially 0$^\circ$ and 180$^\circ$ for orientation I   
and II, respectively. 
We characterize location of the protein by the $z$-coordinate of its
center of mass, $z_{CM}$, and the radial component of the center of mass
in the $x-y$ plane, $r_{CM}$. 
The geometry of proteins is described by the radius of gyration, $R_g$, 
and the vertical thickness, 
$\Delta z=z_{\text{max},C_\alpha}-z_{\text{min},C_\alpha}$, 
estimated based only on the $\alpha-C$ atoms. In addition,
we determine RMSD, the root-mean-square deviation from the native state,
and parameter $w$, constructed from the three eigenvalues
of the moment of inertia. This parameter distinguishes between the flat 
($w<0$), elongated ($w>0$) and globular ($w \approx 0$) shapes  \cite{airwater}.

\section{Results}

\begin{figure}[h]
\centering
\includegraphics[width=0.5\textwidth]{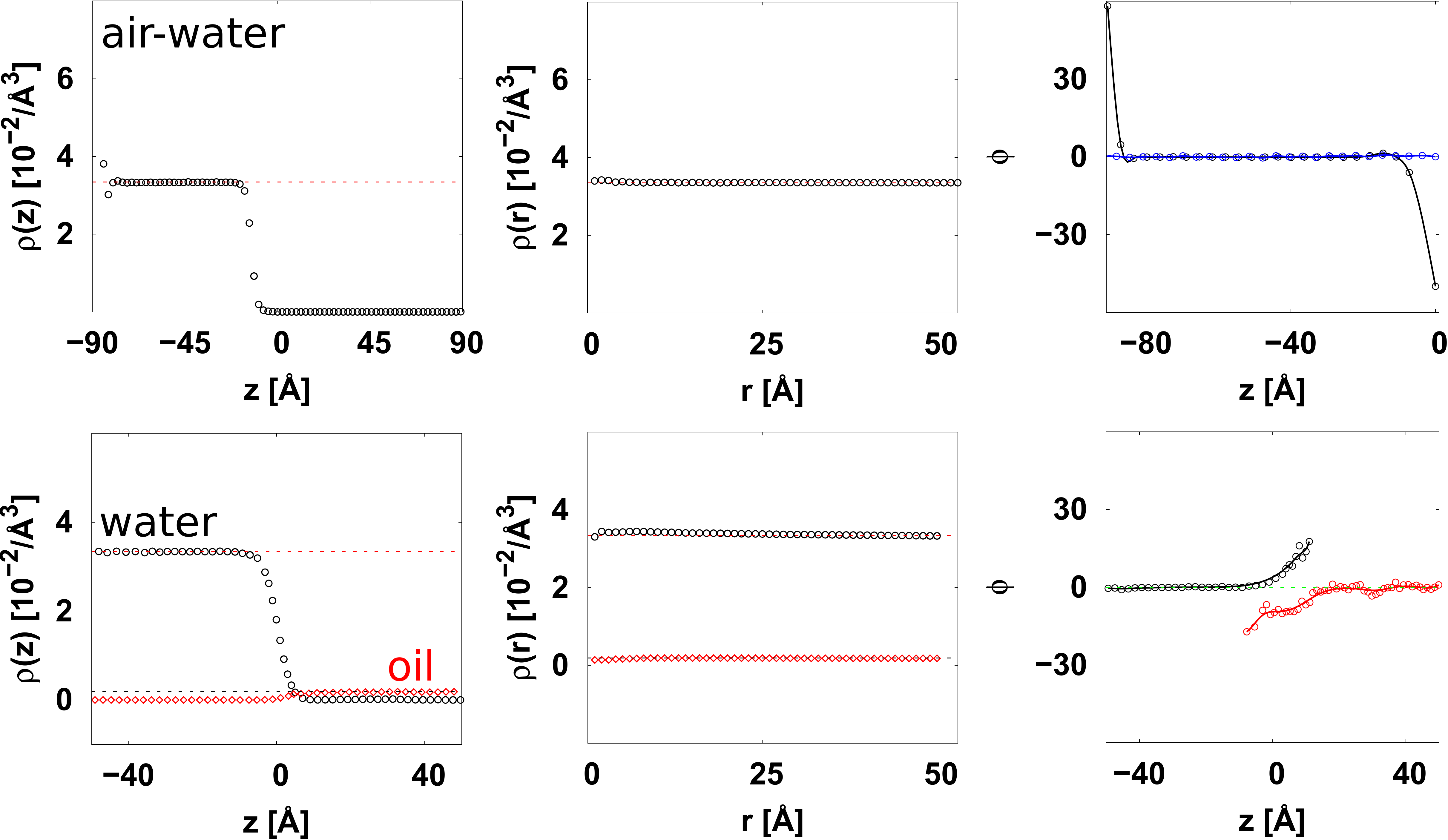}
\caption{The top panels: The number density profiles of water molecules at the air-water
system along $z$-axis and the radial direction in the $x-y$ plane. The red
dashed line indicates the level characterizing water under atmospheric pressure 
in the plots of $\rho(z)$ and $\rho(r)$.
The bottom panels: similar to the top panels but for the oil-water interface.
The red data points show results for the oil molecules.
} \label{densityw}
\end{figure}

On relaxation, the interface becomes rough (some water molecules
escape to the air, creating vapor) and we average the density field over 2 ns.
Fig.~\ref{densityw} shows the number density profile of the water 
molecules along the $z$-axis (away from the wall at the bottom) and along
the radial direction in the $x-y$ plane.
The number density of water in the bulk region
is $3.32 \pm 0.04 \times 10^{-2}$ /{\AA}$^3$,
which is close to $3.34 \times 10^{28}$ m$^{-3}$ for water under the
normal conditions. We observe  that $\rho(z)$ goes down from the bulk value
to zero at $z=-5$ {\AA} and the width of the interface is about 10 {\AA}.
The mid-point of the interface is at --12 {\AA}.
The bulk density of water is
the same at both kinds of the interface.
In the oil-water case, $\rho(z)$ of water goes to zero at 11 {\AA} and the
mid-point of the interface is at 0 {\AA}. 
The density of the
oil molecules disappears at -7 {\AA} and the corresponding
mid-point is at 3 {\AA}.
The width of the interface here can be defined as the separation
between the $z$ values at which water and oil reach their respective
bulk densities. Such a width is about 20 {\AA}, which is twice as
large as the width of the air-water interface.\\

Fig.~\ref{densityw}  (the rightmost panels) also shows the average angle
that the polarization vector of the water molecules makes with the $x-y$ plane.
We observe that, the polarization of the interfacial water 
molecules tends to stay parallel to the interface regardless of the
nature of the interface.\\

\subsection{Proteins at the air-water interface.}

\begin{figure}[h]
\centering
\includegraphics[width=0.8\textwidth]{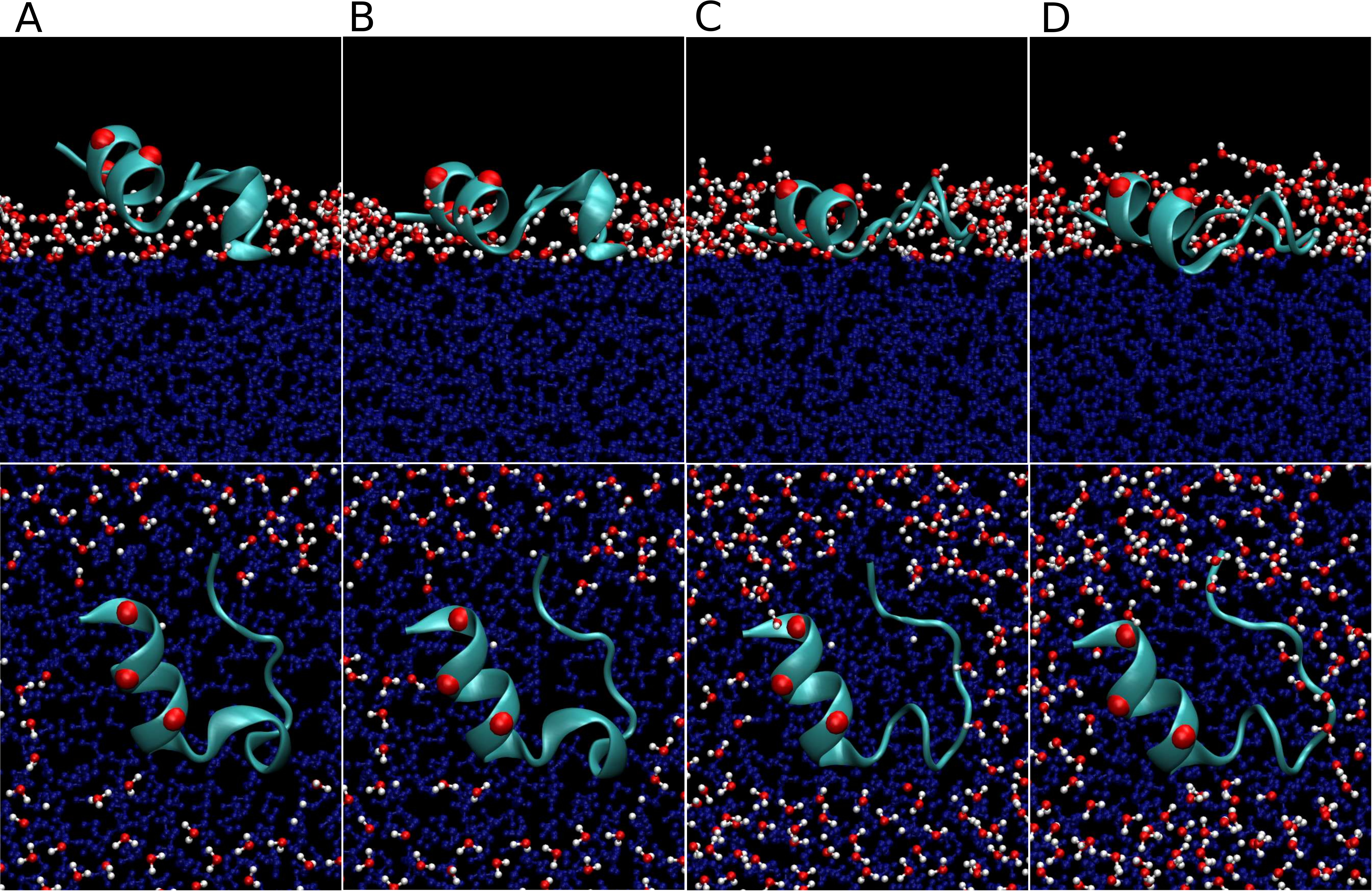}
\caption{Snapshots of protein 1L2Y placed in orientation I. The top 
panels correspond to the side view, and the bottom panels to the top view.
The molecules of water at the interface,  with $z>-12$ {\AA}, 
are shown in red and white and those in the bulk -- in blue.
A: The initial state of the system. B: The state after minimization. 
C: The state after 2 ns equilibration at $T=150$ K. 
D: A state at the end of the 10 ns time evolution at $T=300$ K.
} \label{up1l2y}
\end{figure}

As an illustration, we first consider the short protein 1L2Y
shown in Fig.~\ref{up1l2y} at four stages: the initial
placement in orientation I, after the energy minimization, after
2 ns equilibration at $T$=150 K, and in a snapshot obtained towards the
end of the 10 ns evolution at 300 K.
1L2Y has three hydrophobic residues (shown in red) that are located
in the helical segment at the N-terminus. These residues are
seen to get pinned at the interface whereas the hydrophilic
C-terminal segment moves in water fairly freely, resulting in
sudden changes in the parameters pertaining to the shape,
such as $\Delta z$.\\

\begin{figure}[h]
\centering
\includegraphics[width=0.9\textwidth]{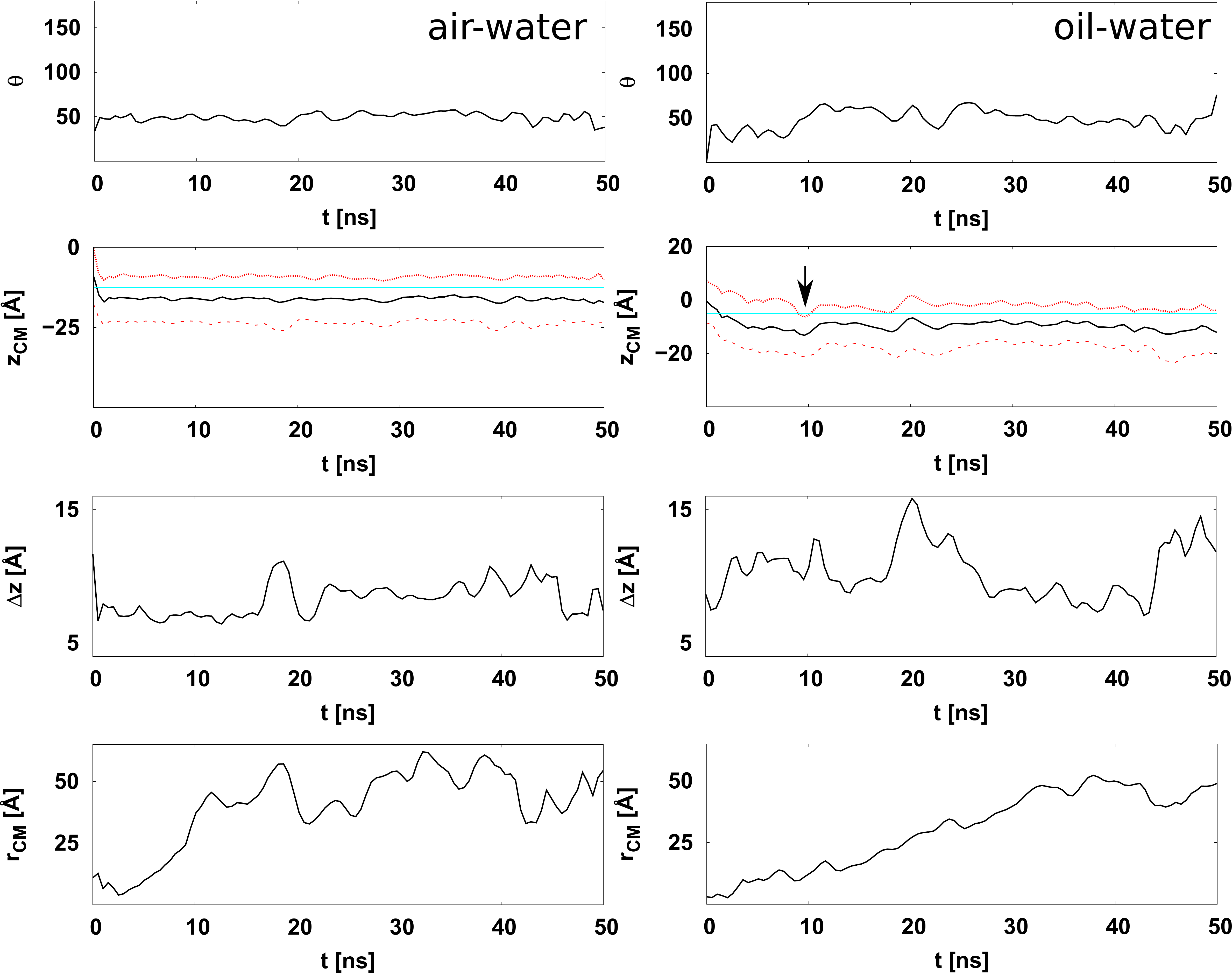}
\caption{Time evolution of $\theta$, $z_{CM}$, $\Delta z$ and $r_{CM}$
of 1L2Y starting in orientation I 
at the air-water (left) and oil-water (right) interfaces at T=300 K. 
In the plot of $z_{CM}$, the black line shows the values of $z_{CM}$, 
the cyan line displays the average position of the interface, 
the red lines above and below the line corresponding to the interface 
indicate the vertical position of the top and bottom atoms of the protein. 
The arrow in the second panel on the right indicates a situation in
which the protein is pinned at the interface but the interface itself is 
below its average position (see the left panel of Fig.~\ref{struoil1lip}).
} \label{t1l2y}
\end{figure}

The left panels of Fig.~\ref{t1l2y} show the time evolution of $\theta$, 
$z_{CM}$, $\Delta z$, and $r_{CM}$ for 1L2Y  at $T$=300 K
when the protein starts in orientation I.
During the 50 ns of the evolution, 1L2Y is seen to stay at the interface
(constant $z_{CM}$) and merely diffusing in the $x-y$ plane (as
evidenced by the behavior of $r_{CM}$).
There are noticeable temporal changes in $\Delta z$ signifying
conformational transformations. For instance,
$\Delta z$ switches from 7.2 to 11.7 {\AA} at $t\sim 17.5$ ns.
The angle $\theta$ is zero initially, but then it stays
more or less fixed at about 50$^\circ$. The fact that this angle
is not 0 reflects two circumstances:
1) the protein gets deformed which changes the direction of $\vec{h}$,
2) the hydropathy indices have not been derived by using
the atomic force field. Nevertheless, the constancy of $\theta$
signifies existence of a preferred orientation, similar to
what is observed in the phenomenological model.\\

\begin{figure}[h]
\centering
\includegraphics[width=0.8\textwidth]{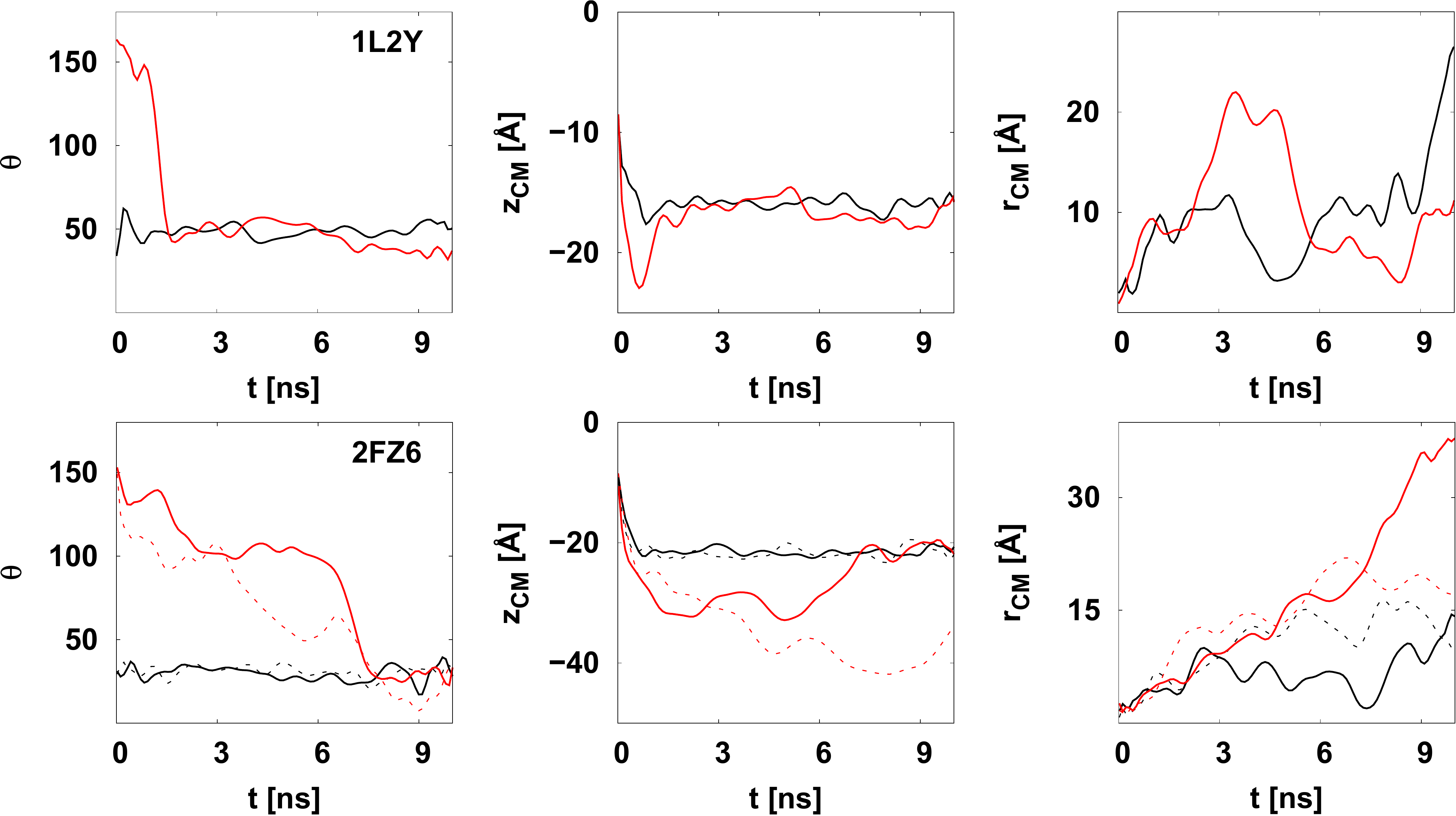}
\caption{Time evolution of $\theta$, $z_{CM}$ and $r_{CM}$ for 1L2Y
(the top panels) and 2FZ6 (the bottom panels) at the air-water interface. 
The black lines are for the starting
orientation I and the red lines are for orientation II. 
For 2FZ6, the solid lines are for $n_{SS}=4$, while the dashed line are for $n_{SS}=0$. 
} \label{type1airw}
\end{figure}

If one starts in orientation II (the top panels of Fig.~\ref{type1airw}),
the protein is seen to first dunk into the water phase completely
where it rotates under the influence of thermal fluctuations.
Subsequently, it diffuses back
to the interface, gets pinned there around 1.4 ns, and $\theta$
comes close to the value obtained for orientation I.\\

A similar behavior is observed for the hydrophobic 2FZ6 (the bottom panels).
There are no desorption events
in orientation I irrespective of the number of the disulfide bonds.
An example of a conformation of 2FZ6 obtained from one trajectory starting in
orientation I is shown in Fig.~S2 in SI.
For orientation II, 2FZ6 gets repinned at 7.1 or 10.1 ns, depending on
whether $n_{SS}$ is 4 or 0 (the latter is beyond the scale of the figure).
The reduction in $n_{SS}$ appears to slow the rotation of the
protein which delays adsorption to the interface.\\

\begin{figure}[h]
\centering
\includegraphics[width=0.9\textwidth]{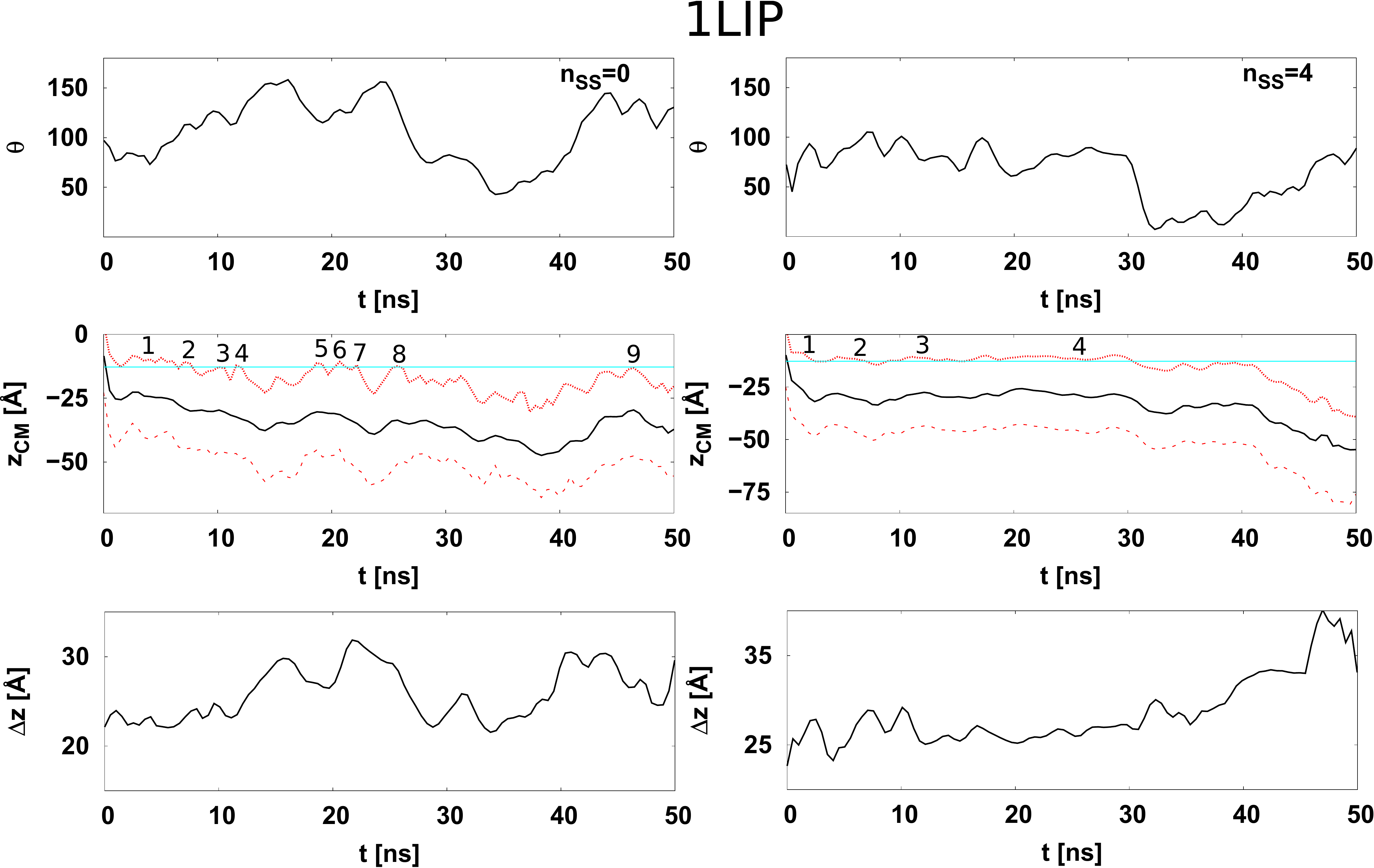}
\caption{Time evolution of $\theta$, $z_{CM}$ and $\Delta z$ of 1LIP at $n_{SS}=0$ (left) in orientation II 
and $n_{SS}=4$ (right) in orientation I at the air-water interface.
In the plot of $z_{CM}$, the black line shows the values of $z_{CM}$, the cyan line displays the average position of the air-water 
interface, the red lines above and below the interface are the top and bottom atoms of the protein along $z$-axis. 
} \label{t1lipx0}
\end{figure}

The other three proteins, 1LIP, 1GB1 and 2LYZ, are overall hydrophilic, but
unlike 1L2Y, their hydrophobic residues are distributed in a dispersed manner.
These proteins can adsorb to the interface and then leave it
multiply, as illustrated in Fig.~\ref{t1lipx0} for 1LIP.
The left panels are for $n_{SS}$=0 and orientation II. Nine absorption
events are seen. The first of them starts at the begining
and it lasts for 6.34 ns (the time counts from the last
equilibrated conformation obtained at 150 K). The protein then dips
into the water phase for 0.3 ns and afterwords it gets pinned
for 1.68 ns. The durations of the remaining adsorption events
observed are: 0.06, 0.28, 0.88, 0.72, 0.20, 1.00 and 0.12 ns successively.
Each adsorption event comes with its own type of deformation, as
evidenced by the different corresponding values of $\Delta z$.
The right panels of Fig.~\ref{t1lipx0} are for $n_{SS}$=4
and orientation I. In this case, we observe four adsorption
events. They last for 2.4, 2.8, 4.6 and 14.3 ns successively.
The statistics of the adsorption times are too small to assess
but they suggest a uniform distribution except for the regime
of the short lasting events which corresponds to a border-line
behavior. \\

\begin{figure}[h]
\centering
\includegraphics[width=0.8\textwidth]{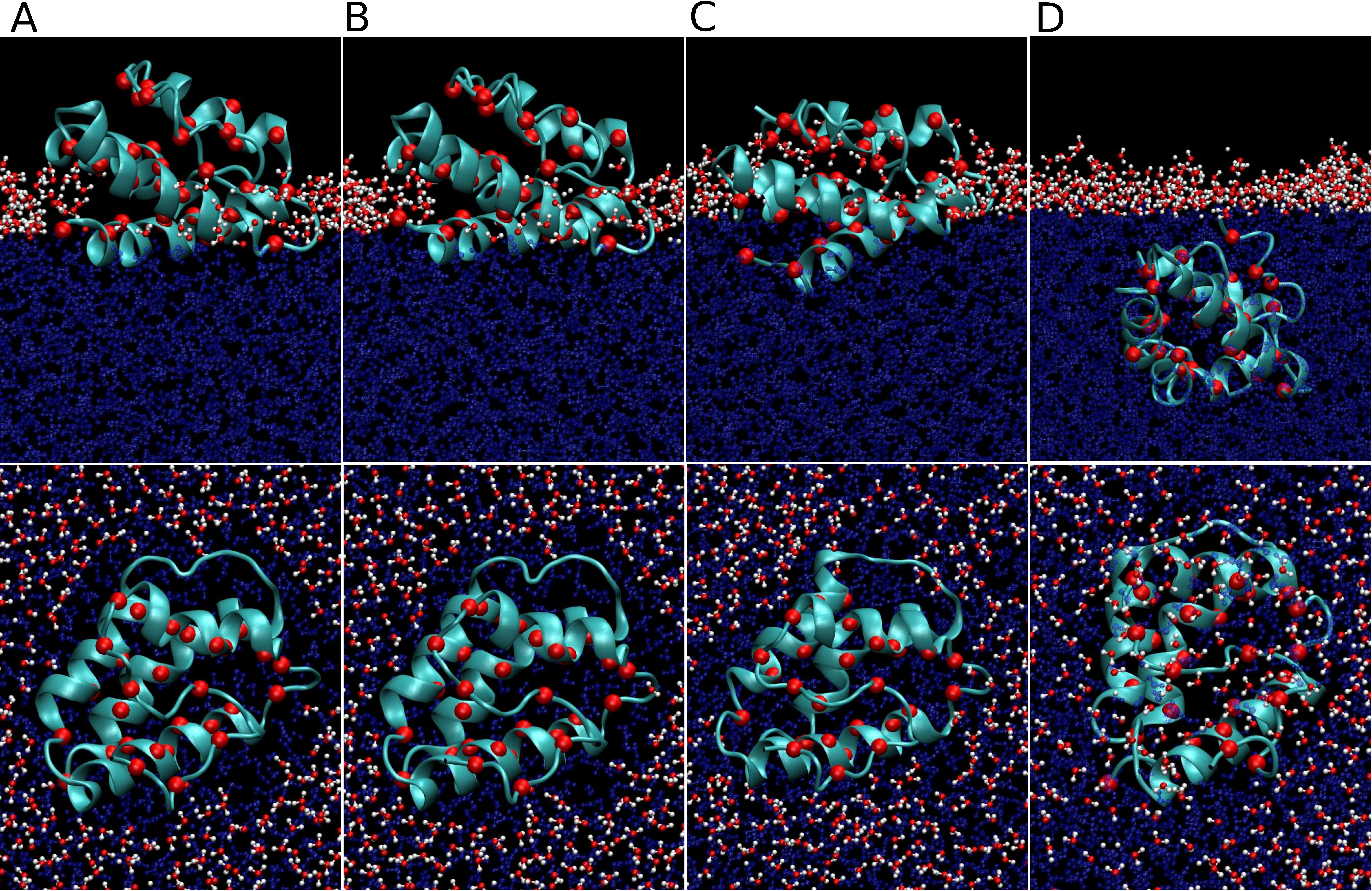}
\caption{Protein 1LIP with $n_{SS}=0$ inityially placed in orientation II. 
A: The initial state of the system. B: The state after the energy minimization. 
C: The state after 2 ns equilibration at $T=150$ K. D: 
The final state of the system after 10 ns equilibration at $T=300$ K.
The top panels show the side view, and the bottom panels -- the top view.
} \label{do1lip}
\end{figure}

\begin{figure}[h]
\centering
\includegraphics[width=0.6\textwidth]{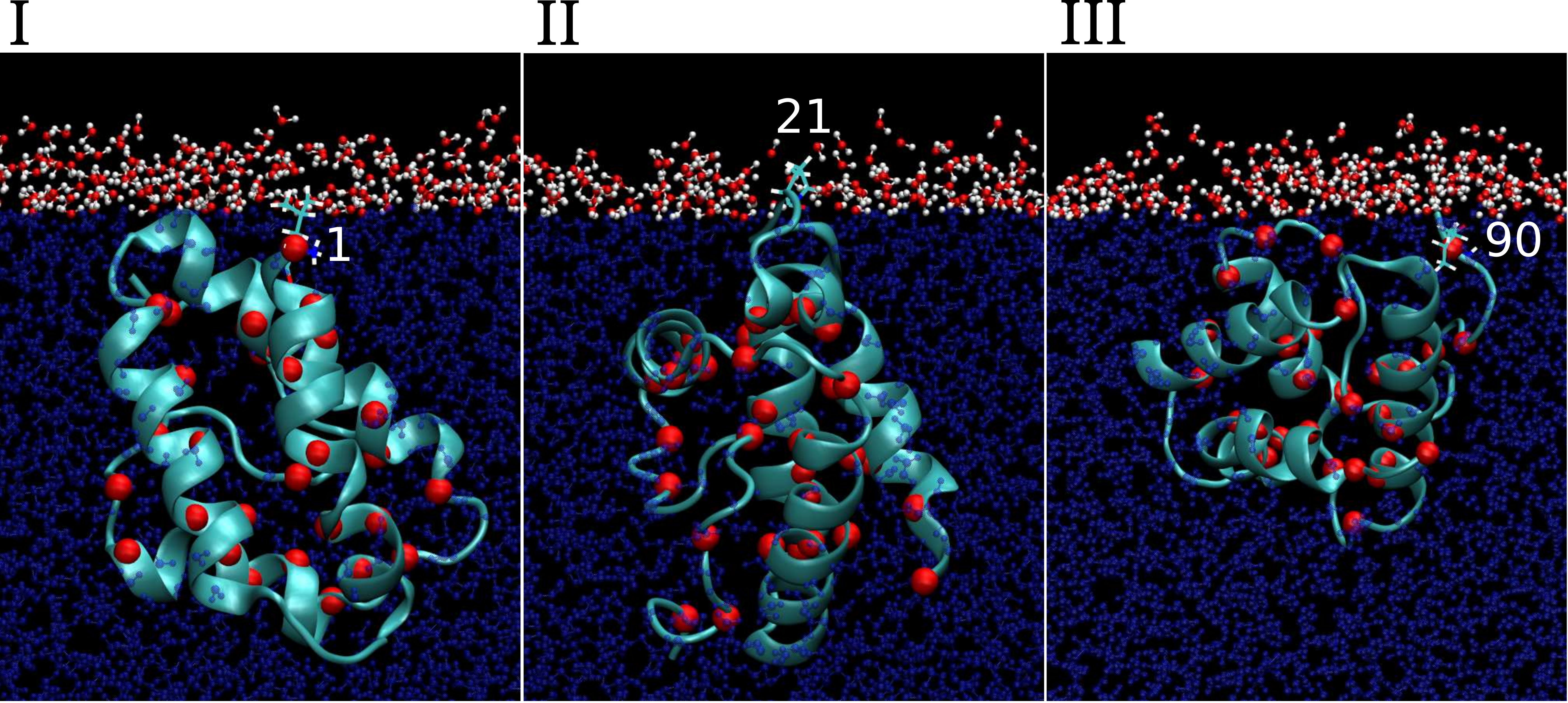}
\caption{Examples of conformations of 1LIP adsorbed at the
air-water interface. 
In panels I through III, pinning is initiated by ILE-1 
(for $n_{SS}$=4), PRO-21 (for $n_{SS}$=0), and ILE-90 
(for $n_{SS}$=0) respectively. 
} \label{so1lip}
\end{figure}

The pinning of 1LIP to the interface is usually driven by just one
residue. In the left panels of Fig.~\ref{t1lipx0}, i.e. for $n_{SS}=0$,
it is ILE-90 in all nine events. One corresponding
conformation is shown in Fig. \ref{do1lip}.
For $n_{SS}$=4, illustrated in Fig. \ref{so1lip},
the protein is more compact and observe more variety.
The left panel shows pinning by LEU-1, which is observed in
70\% of the trajectories. The middle panel shows pinning by PRO-21
and the right panel -- by ILE-90. We have also observed pinning
by LEU-63. All of these residues are hydrophobic except for
PRO which is moderately hydrophilic in the Kyte and Doolittle
scale ($q_i$=-1.6). However, in the CHARMM22 force-field, PRO
appears to behave as a hydrophobic residue. The special
character of PRO is discussed in ref.~\cite{Palliser}. Various
values of the measured hydrophobicity of PRO have been reported. 
In particular Table I in ref. ~\cite{Palliser} cites the value of 2.34.\\


Proteins 1GB1 and 2LYZ  are more hydrophilic than 1LIP so they are
expected to be able to leave the interface with a greater ease.
Nevertheless, they do get pinned. Fig.~S3 in SI shows
an example of pinning by VAL-21 in 1GB1 (the left panels) and
by PRO-70 in 2LYZ with $n_{SS}$=4 (the right panels).
If the trajectories were starting from orientation I,
the pinning events lasted for 0.06 and 0.08 ns for 1GB1 and
2LYZ respectively (0.10 ns for 2LYZ if $n_{SS}$=0). 
In 5 trajectories of 10 ns, we observe
no return to the interface in either of the two proteins. However,
if one uses the aligned crystal structure refers to the optimal surface 
orientation in the initial state (named orientation O), the average duration time 
at the interface at T=300 K is 6.2 ns for 1GB1 and 3.4 (5.5) ns for 2LYZ with 
$n_{SS}=4$ ($n_{SS}=0$). This indicates that, proper surface
orientation plays a key role in the adsorption of proteins at the interface.\\

It should be noted that our results pertain to single proteins.
Their depinning will be inhibited, but not eliminated, if it is
a part of a dense film \cite{airwater}.

\subsection{Proteins at the oil-water interface.}

\begin{figure}[h]
\centering
\includegraphics[width=0.7\textwidth]{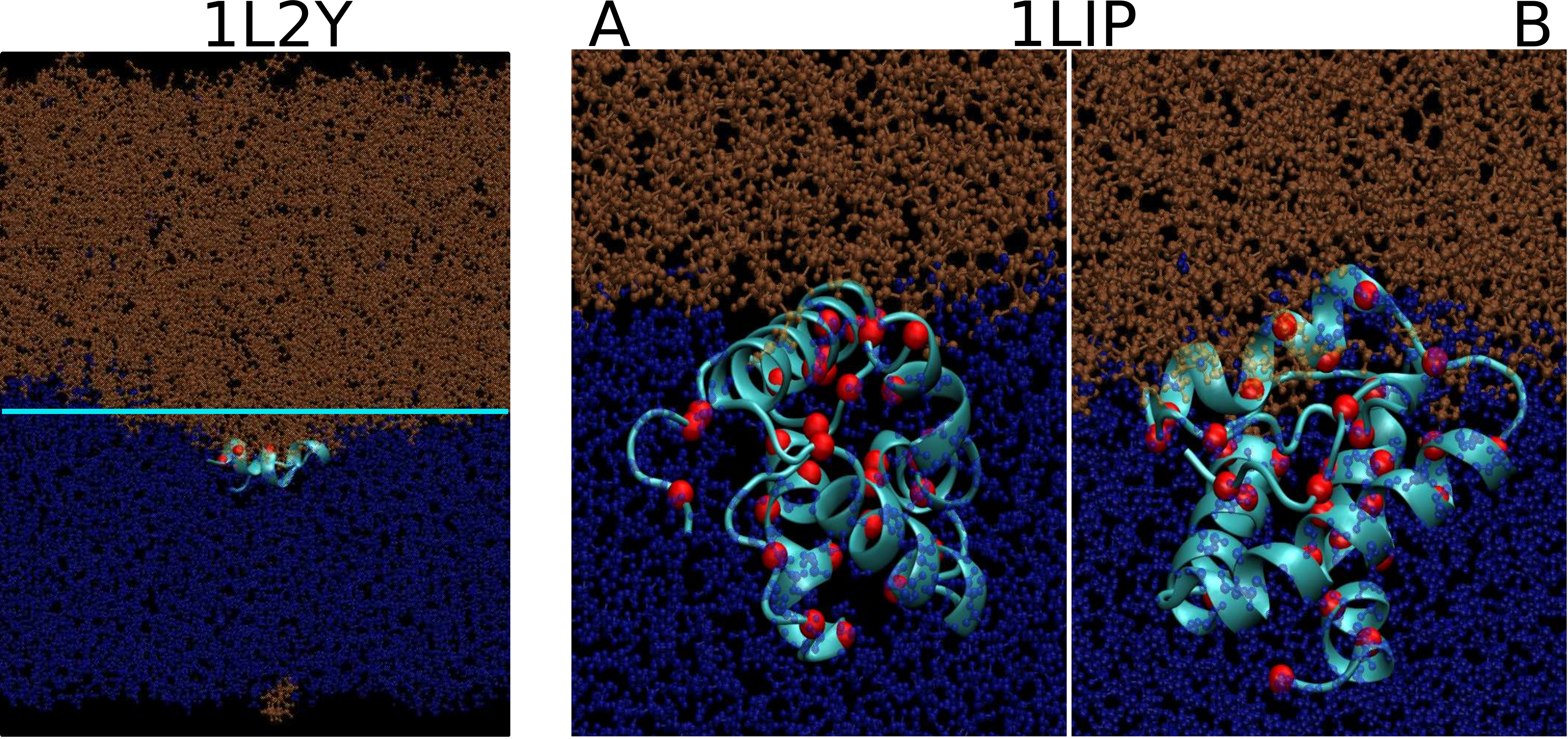}
\caption{Left: A snapshot of 1L2Y adsorbed at the oil-water interface. 
The cyan line indicates the average position of the interface. 
The molecules of oil are shown in orange. 
Right: 1LIP adsorbed at the oil-water interface at $T=300$ K.  
A: 1LIP with $n_{SS}=4$ after starting in orientation I. 
B: 1LIP with $n_{SS}=0$ after starting in orientation II.
} \label{struoil1lip}
\end{figure}

The oil-water interface is broader and rougher than the air-water one.
As a result, pinning is driven not by single residues but by more 
extended regions containing hydrophobic residues and the tendency to 
stay at the interface is generally stronger. 
An example of pinning of 1L2Y is shown in the left panel of 
Fig.~\ref{struoil1lip}.
In this example, the protein is seen to adhere to the interface 
along its length, but further evolution generates conformations
which partially stretch away from the interface. 
The right-hand panels of
Fig.~\ref{t1l2y} show the corresponding evolution of $\theta$,
$z_{CM}$, $\Delta z$ and $r_{CM}$ that starts from orientation I.
The angle $\theta$ is close to the one at the air-water interface,
but it displays bigger fluctuations. 
However, the variations in $\Delta z$ are stronger.
The plot of $z_{CM}$ suggests that the protein dips down from the
average location of the middle of the interface (as indicated by the
arrow) but, in fact, it stays pinned to the interface because
the interface itself exhibits a change in shape.
Also protein 2FZ6 stays at the interface for $n_{SS}$ of 4 and 0
(not shown), which is analogous to what we observed for the air-water case.\\

Snapshots of the pinned conformations of 1LIP
are shown in Fig.~\ref{struoil1lip}.
The left panel corresponds to $n_{SS}=4$ and the pinning
occurs at residues 1--4 (LEU, ASN, CYS, GLY)
and 40--47 (SER, SER, GLY, ASP, ARG, GLN, THR, VAL).
The right panel is for $n_{SS}=0$ and the pinning involves regions
17--23 (VAL, GLN, GLY, GLY, PRO, GLY, PRO), 
61--73 (LEU, ASN, LEU, ASN, ASN, ALA, ALA, SER, ILE, PRO, SER, LYS, CYS),
and 82 to 85 (SER, PRO, ASP, ILE). Such extended pinning 
regions are not observed in the case of air-water interface.
The difference in behavior stems from the fact that the
air-water interface is better defined, and thus corresponding to a
larger surface tension, than the intermixed oil-water interface.
The larger surface tension means that the protein tends to
minimize its exposure to the air.\\

\begin{figure}[h]
\centering
\includegraphics[width=0.9\textwidth]{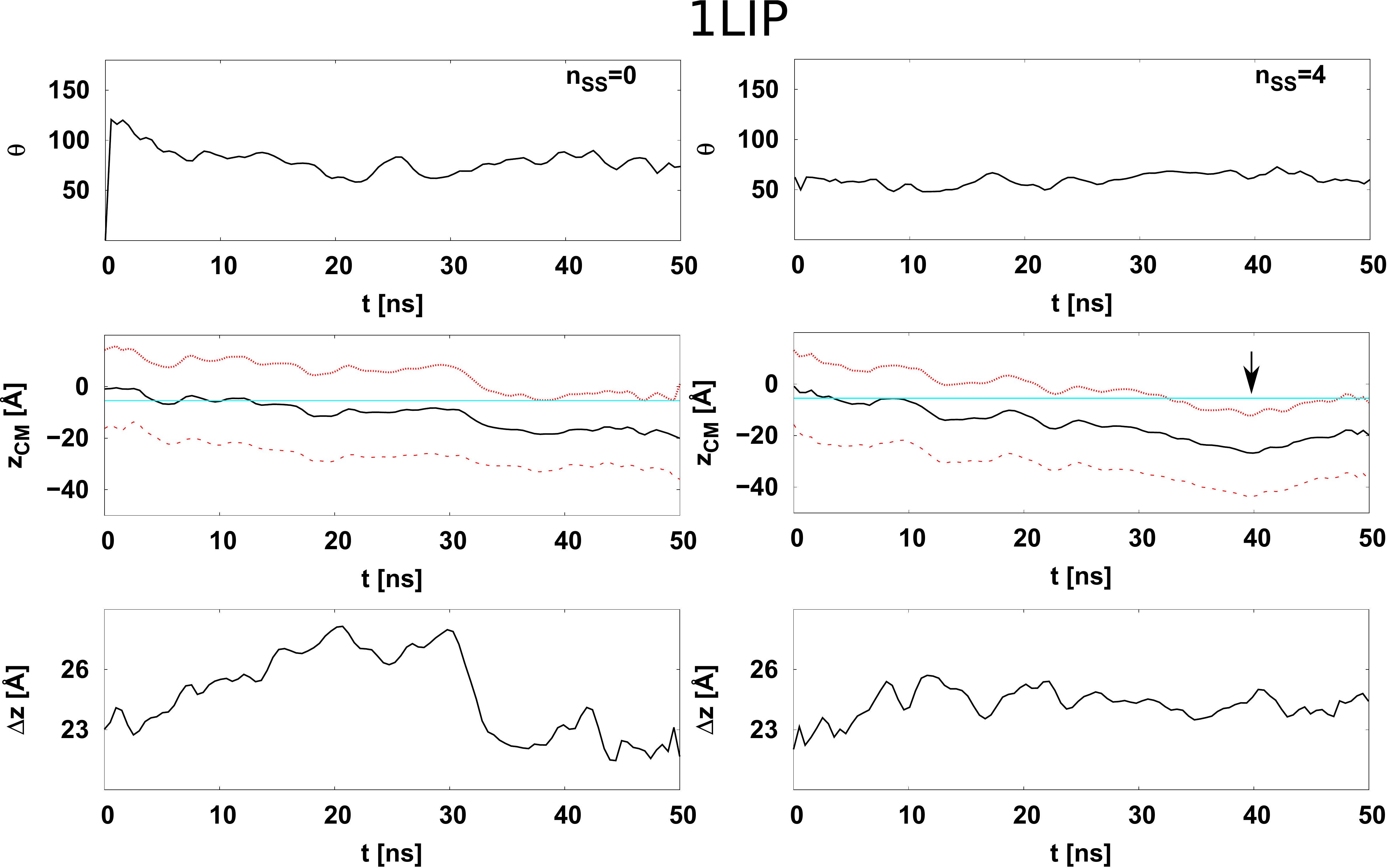}
\caption{Similar to Fig.~\ref{t1lipx0}, but for the oil-water interface.
} \label{poil1lip}
\end{figure}

 
Fig.~\ref{poil1lip} shows that this protein does not depin during
the simulations for both considered values of $n_{SS}$ indicating
a stronger coupling to the interface compared to the air-water case.
$\Delta z$
undergoes somewhat larger fluctuations for $n_{SS}$=0 than for
$n_{SS}$=4 and the protein with all disulfide bonds present
stays deeper in water.
Fig.~S4 in SI demonstrates that the shape-related parameters
$R_g$, RMSD and $w$ of 1L2Y and 1LIP  at the oil-water interface
are similar to those at the air-water one. 
The same conclusion holds true for proteins 1GB1 and 2LYZ, i. e. they
depin from the air-water interface, but not from the oil-water interface.
These results suggest that, the strength of pinning a protein may vary 
significantly  at different interfaces but its interfacial-related distortion is 
comparable. For example, the average RMSD of 1GB1 is 1.15 {\AA} 
when it is adsorbed at 
the air-water interface, while it is 1.20 {\AA} at the oil-water interface.
The amplitude of the distortion in all-atom simulations is roughly 7 times smaller 
than that presented in ref. \cite{airwater} for the same protein (see Fig. 2 of ref. 
\cite{airwater}). Nevertheless, one has to note that, the strength of the 
interface-related force is selected as $A=10~\epsilon$ in 
the coarse-grained model to 
prevent the depinning of proteins. The distortion of proteins can be 
reduced by decreasing the values of $A$. For instance, the average RMSD of 
1GB1 at the interface decreases from 8.80 {\AA} at $A=10~\epsilon$ to 7.30 {\AA} at 
$A=5~\epsilon$. Moreover, the possibility of depinning from the interface 
increases as $A$ decreases in the coarse-grained model. 
Another primary characteristic
of the distortion of proteins is $\Delta z$, its values obtained in the all-atom
simulations agree with that in the coarse-grained model (see Fig. 13 of 
ref.~\cite{beer} for 1LIP). 
\\

It is interesting to determine the two-dimensional diffusion coefficient
$D_2=\frac{\langle\Delta \vec{r}\rangle ^2}{4\Delta t}$ of a protein at the interface.
Here, $\Delta \vec{r}$ is the displacement of the protein determined over each
lag time $\Delta t$. 
$D_2$ is determined in the asymptotic time regime.
In the coarse-grained model \cite{airwater}, and in the limit of
a dilute protein film, one gets 0.0630 and 0.0245 {\AA}$^2$/$\tau$
for 1GB1 and 2LYZ respectively. The time scale, $\tau$, is of order
1 ns and the results were obtained by considering trajectories
lasting for 150 000 $\tau$ and by averaging over many proteins 
when at the interface. The time scale of the all-atom simulation
of 10 ns is too short to approach the necessary asymptotic behavior
to obtain the saturation in the effective time-dependent $D_2$.
We can only investigate the initial dependence of such a $D_2$.
We measure $\Delta \vec{r}$  every $\Delta t=0.02$ ns and 
averaged over 10 trajectories. Generally, we get the
effective $D_2$ to be substantially larger (typically by a 
factor of 50). Other than the short time scales 
(the lag time in the all-atom simulations is 0.02 ns but
1 ns in the coarse-grained model), the reason
for the larger value is that in all-atom model, thermal fluctuations
affect all heavy atoms, and not only the $\alpha$-C atoms.\\

For the oil-water interface, the short time values of 
$D_2$ are $11.74\pm 0.6$, $5.62\pm 0.3$, 
$5.14\pm 0.3$, $6.65\pm 0.5$ and $3.73\pm 0.2$ {\AA}$^2$/ns for 1L2Y, 
2FZ6 ($n_{SS}=4$), 1LIP ($n_{SS}=4$), 1GB1 and 2LYZ ($n_{SS}=4$), respectively. 
These results show that $D_2$ depends primarily on the sequential
length of the proteins: the longer the protein the slower the diffusion. 
For the air-water interface,  we get $D_2$ of 27.74$\pm 1.8$ and $10.54\pm 0.4$
{\AA}$^2$/ns for 1L2Y and 2FZ6 ($n_{SS}=4$) respectively. Thus the
diffusion at the air-water interface appears to be about twice as fast
as at the oil-water one. This is because the oil phase anchors more
hydrophobic residues. We also observe that the reduction
of the disulfide bonds results in slowing the diffusion down. For
instance, 2FZ6 with $n_{SS}=0$ has $D_2$ of $D=9.77\pm 0.2$ nm$^2$/ns at
the air-water interface, which is about  7\% smaller than when $n_{SS}=4$.\\

LTP1 has a post-translationally modified isoform LTP1b
(PDB: 3GSH), which contains a fatty ligand ASY which stands for
$\alpha$-ketol, 9-hydroxy-10-oxo-12($Z$)-octadecenoic acid \cite{bakan2009}.
The C9 carbon of this ligand is covalently bound to the O2 
atom of Asp-7 of the protein. The bonding site partitions the 
ligand ASY into two branches. The one from C10 to C18 is buried 
inside the cavity of LTP1b, while another branch from C1 to C8  
lies on the surface of the protein and is hydrophobic. 
In the coarse-grained model \cite{beer}, 
the presence of the ligand contributes to a better adsorption of LTP1b
to the air-water interface compared to LTP1 \cite{beer} as ligand
stretches out of the cavity. \\

\begin{figure}[h]
\centering
\includegraphics[width=0.9\textwidth]{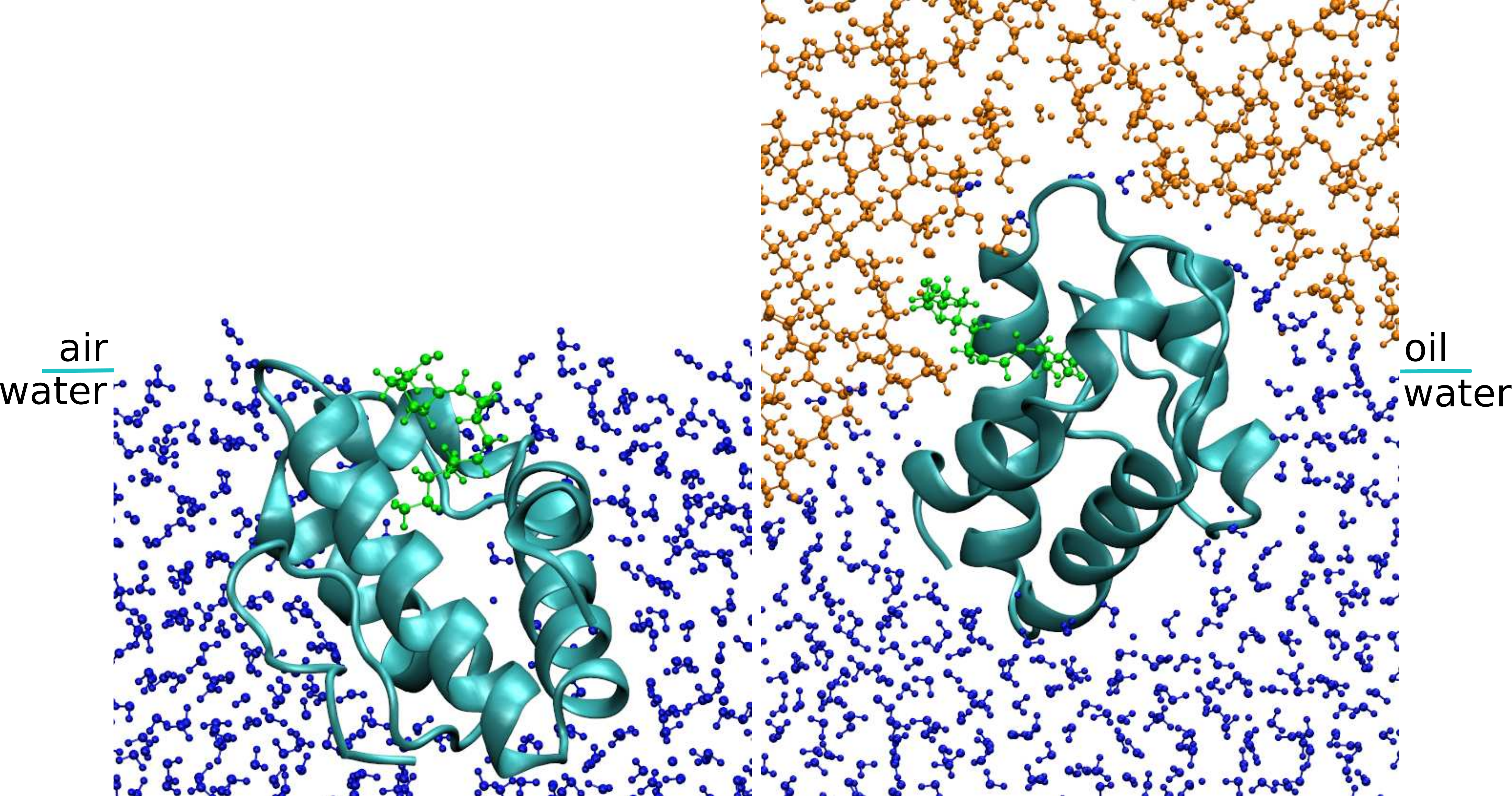}
\caption{Side view of LTP1b at the air-water (left panel) and
oil-water (right panel) interfaces. The atoms of the ligand are shown in green.
The water molecules are in blue and those of oil in orange.
} \label{ligand}
\end{figure}

In order to compare the behavior of the ligand at all-atom level to the
coarse-grained model, we simulate LTP1b ($n_{SS}=4$) with orientation I 
at both the air-water and oil-water 
interfaces. The system is evolved at $T=300$ K for 10 ns.
At the air-water interface, LTP1b can be adsorbed to the interface and then 
depinned multiple times just like LTP1. The branch of the ligand 
from C10 to C18 remains in the cavity throughout the simulation. 
The pinning of LTP1b can take place either at the C1 and C2 atoms
of the ligand, as illustrated in the left panel of Fig.~\ref{ligand},
or at PRO-21 of the protein. Both pinning centers
are then exposed to the air.
In the case of the oil-water interface, the buried branch of ligand still 
stays inside the cavity, but the exposed branch is fully in contact with the
oil molecules, as shown in the right panel of Fig.~\ref{ligand}.
As a result, there is no depinning, highlighting
the differences between the two interfaces.

\section{Conclusions}

The differences in the behavior of proteins placed at the the air-water
and oil-water interfaces have been attested experimentally
by Sengupta and Damodaran \cite{Sengupta}.
They have found that the rates of adsorption of $\beta$-casein, BSA,
and lysozyme to the oil-water interface were higher by an order
of magnitude compared to the air-water interface. The oil used
was triolein. They attributed the differences to the nature of the
dispersive interactions at the interface: attractive in the case of
oil and repulsive in the case of air.\\

Even though the two interfaces have different physical
properties, the interfacial behavior of the five proteins studied here
is fairly similar: there is a distortion and pinning.
Thus the phenomenological coarse-grained model based on the 
hydropathy indices appears to capture the physics of pinning and distortion
for both types of the interfaces. 
The model involves the
amplitude of the hydropathy force and 
weakening of this amplitude can account for the depinning
situations. However, some proteins are seen to depin and some
do not, at least during the duration of the simulations. 
Thus the adjustment of the amplitude should be done
on a case by case basis. In the oil-water case, our original
amplitude $A=10\;\epsilon$ would be then adequate in this respect.\\

It would be interesting to work out some rules that
predict which hydrophilic proteins can depin and which can not. 
These rules should depend on the type of the interface.
They should also reflect the observation that the hydrophobic
residues appear to interact with the oil stronger 
than with the air (which, in our model, is represented simply
by the absence of molecules).
This effect can be captured by some strengthening of the
Kyte and Doolittle \cite{Kyte} hydropathy indices for the
hydrophobic residues when dealing with oil.\\

These issues could perhaps be sorted out by considering artificial
systems in which proteins are replaced by ligand-grafted nanoparticles 
as considered by Garbin et al. \cite{Garbin}. The ligand used
in this case was amphiphilic mercapto-undecyl-tetra (ethylene glycol)
and the experiment involved measurements of the surface pressure.
One may consider grafting chains with homopeptidic tails to the
nanoparticles to quantify interactions with the specific residues.

{\bf Acknowledgements}

This research has been supported by the National Science Centre, Poland,
under grant No.~2014/15/B/ST3/01905 and by the EU
Joint Programme in Neurodegenerative Diseases project (JPND CD FP-688-059) 
through the National Science Centre (2014/15/Z/NZ1/00037) in Poland.
The computer resources were supported by the PL-GRID infrastructure
and financed by the European Regional Development Fund under the
Operational Programme Innovative Economy NanoFun POIG.02.02.00-00-025/09.

\clearpage

\end{document}